\documentclass[10pt,twocolumn,letterpaper]{article}

\usepackage{cvpr}
\usepackage{times}
\usepackage{epsfig}
\usepackage{graphicx}
\usepackage{amsmath}
\usepackage{amssymb}
\usepackage{xcolor}
\usepackage{multirow}
\usepackage{multicol}
\usepackage{subfigure, enumitem}
\usepackage{threeparttable}
\usepackage[breaklinks=true,bookmarks=false]{hyperref}

\cvprfinalcopy % *** Uncomment this line for the final submission

\def\cvprPaperID{****} % *** Enter the CVPR Paper ID here

\begin{document}

%%%%%%%%% TITLE
\title{Learning for Video Compression with Hierarchical Quality \\ and Recurrent Enhancement}

\author{Ren Yang\\{\small ren.yang@vision.ee.ethz.ch}
\and
Fabian Mentzer\\{\small mentzerf@vision.ee.ethz.ch}
\and
Luc Van Gool\\{\small vangool@vision.ee.ethz.ch}
\and
Radu Timofte\\{\small timofter@vision.ee.ethz.ch}
\and
ETH Z\"urich, Switzerland
}

\maketitle

\begin{abstract}
In this paper, we propose a Hierarchical Learned Video Compression (HLVC) method with three hierarchical quality layers and a recurrent enhancement network. The frames in the first layer are compressed by an image compression method with the highest quality. Using these frames as references, we propose the Bi-Directional Deep Compression (BDDC) network to compress the second layer with relatively high quality. Then, the third layer frames are compressed with the lowest quality, by the proposed Single Motion Deep Compression (SMDC) network, which adopts a single motion map to estimate the motions of multiple frames, thus saving bits for motion information. In our deep decoder, we develop the Weighted Recurrent Quality Enhancement (WRQE) network, which takes both compressed frames and the bit stream as inputs. In the recurrent cell of WRQE, the memory and update signal are weighted by quality features to reasonably leverage multi-frame information for enhancement. In our HLVC approach, the hierarchical quality benefits the coding efficiency, since the high quality information facilitates the compression and enhancement of low quality frames at encoder and decoder sides, respectively. Finally, the experiments validate that our HLVC approach advances the state-of-the-art of deep video compression methods, and outperforms the ``Low-Delay P (LDP) very fast'' mode of x265 in terms of both PSNR and MS-SSIM.
The project page is at \url{https://github.com/RenYang-home/HLVC}.
\end{abstract}

%%%%%%%%% BODY TEXT

\section{Introduction}

In recent years, video streaming over the Internet has become more and more popular. According to the Cisco Forecast~\cite{Cisco}, video generates 70\% to 80\% traffic of mobile data. The proportion of high resolution video is also rapidly increasing.
To be able to more efficiently transmit high quality videos over the bandwidth-limited Internet, it is necessary to improve the performance of video compression.
During the past decades, plenty of video compression standards were proposed, such as H.264~\cite{wiegand2003overview}, H.265~\cite{sullivan2012overview}, \etc. However, these traditional codecs are handcrafted and cannot be optimized in an end-to-end manner.

Recent studies in learned image compression, \eg,~\cite{balle2017end, balle2018variational}, show the great potential of deep learning for improving the rate-distortion performance. It is therefore not surprising to see increasing interest in compressing video with Deep Neural Networks (DNNs) \cite{chen2019learning,wu2018video,cheng2019learning, lu2019dvc,habibian2019video}. For example, Lu~\etal~\cite{lu2019dvc} proposed using optical flow for motion compensation and applying auto-encoders to compress the flow and residual. Then, Habibian \etal~\cite{habibian2019video} proposed a 3D auto-encoder for video compression with an autoregressive prior. In these methods, the models are trained with one loss function and applied on all frames. As such, they fail to generate hierarchical quality layers, in which high quality frames are beneficial for the compression and the post-processing of other frames.

\begin{figure}[!t]
\begin{center}
\includegraphics[width=1\linewidth]{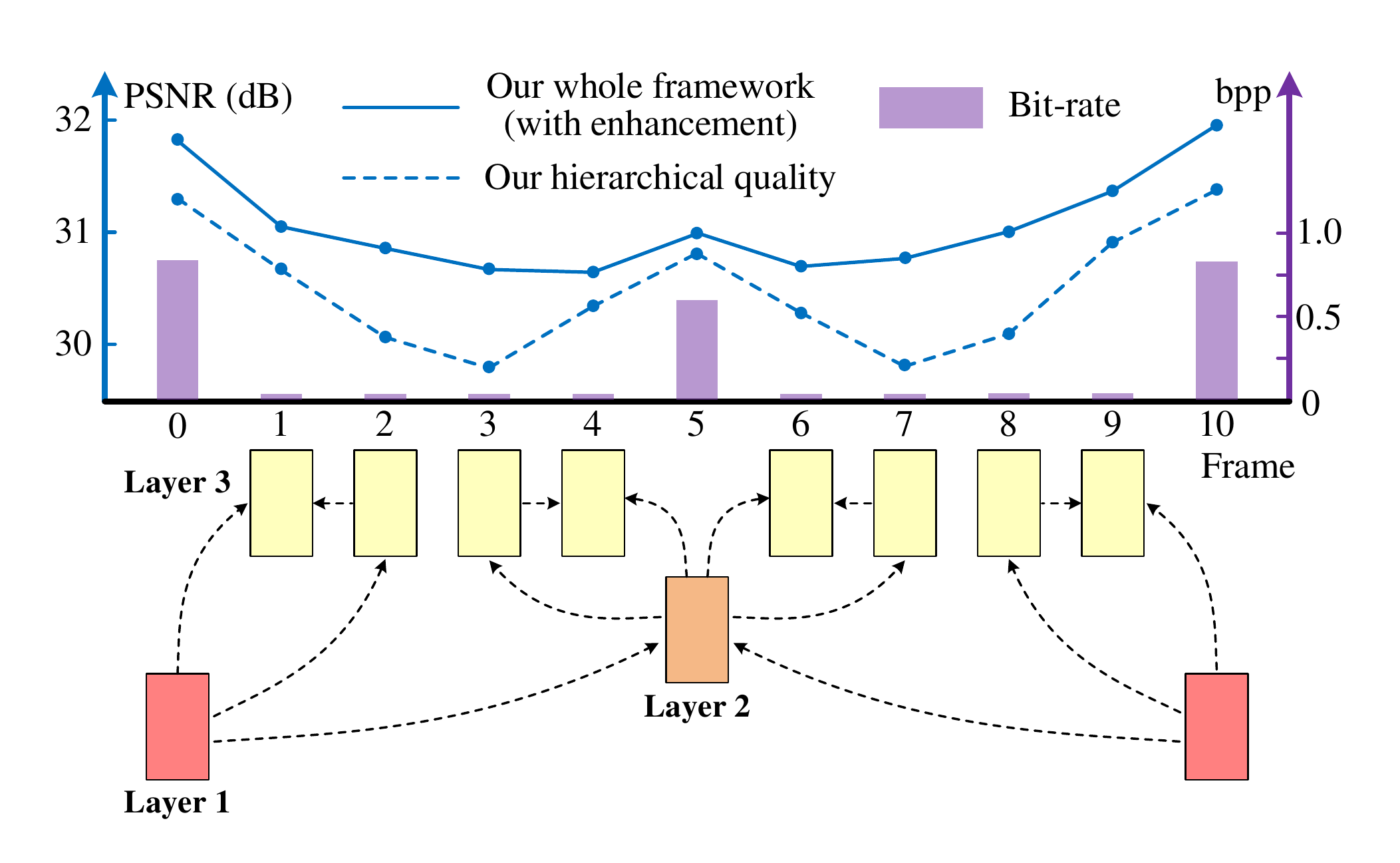}
\end{center}
\caption{The hierarchical layers and the rate-distortion performance in each layer of our HLVC approach. We use the first Group Of Picture (GOP) in the sequence \textit{BlowingBubbles} as an example.
}\label{fig:1}

\end{figure}

This paper proposes a Hierarchical Learned Video Compression (HLVC) method with three hierarchical quality layers and a recurrent enhancement network. As illustrated in Figure~\ref{fig:1}, the frames in layers 1, 2 and 3 are compressed with the highest, medium and the lowest quality, respectively. The benefits of hierarchical quality are two-fold: First, the high quality frames, which provide high quality references, are able to improve the compression performance of other frames at the encoder side; Second, because of the high correlation among neighboring frames, at the decoder side, the low quality frames can be enhanced by making use of the advantageous information in high quality frames. The enhancement improves quality without bit-rate overhead, thus improving the rate-distortion performance. For example, the frames 3 and 8 in Figure~\ref{fig:1}, which belong to layer 3, are compressed with low quality and bit-rate. Then, our recurrent enhancement network significantly improves their quality, taking advantage of higher quality frames, \eg, frames 0 and 5. As a result, the frames 3 and 8 reach comparable quality to frame 5 in layer 2, but consume much less bit-rate. Therefore, our HLVC approach achieves efficient video compression.

In our HLVC approach, we use image compression method to compression layer 1. For layer 2, we propose the Bi-Directional Deep Compression (BDDC) network, which uses the compressed frames of layer 1 as bi-directional references. Then, because of the correlation between motions of neighboring frames, we propose compressing layer 3 by our Single Motion Deep Compression (SMDC) network. The SMDC network applies a single motion map to estimate motions among several frames to reduce the bit-rate for encoding motion maps. Finally, we develop the Weighted Recurrent Quality Enhancement (WRQE) network based on~\cite{yang2019quality}, in which the recurrent cells are weighted by quality features to reasonably apply multi-frame information for recurrent enhancement. The experiments show that our HLVC approach achieves the state-of-the-art performance in learned video compression methods, and outperforms x265's ``Low-Delay P (LDP) very fast'' mode. Moreover, the ablation studies prove the effectiveness of each network in our approach.

\section{Related works}

\textbf{Deep image compression.} In the past decades, plenty of handcrafted image compression standards were proposed, such as JPEG~\cite{wallace1992jpeg}, JPEG 2000~\cite{skodras2001jpeg} and BPG~\cite{BPG}.
Recently,
DNNs have also been successfully applied to improve the performance of image compression~\cite{Toderici2016Variable,toderici2017full,agustsson2017soft,theis2017lossy, balle2017end,balle2018variational,minnen2018joint,mentzer2018conditional,li2018learning,johnston2018improved,lee2019context}. Ball{\'e}~\etal~\cite{balle2017end, balle2018variational} proposed various end-to-end DNN frameworks for image compression, applying the factorized-prior~\cite{balle2017end} and hyperprior~\cite{balle2018variational} density models to estimate entropy. Later, hierarchical prior~\cite{minnen2018joint} and context-adaptive~\cite{lee2019context} entropy models were designed to further advance the rate-distortion performance, and they outperform the state-of-the-art traditional image codec.
Moreover, recurrent structures are also adopted in image compression networks~\cite{Toderici2016Variable, toderici2017full, johnston2018improved}.

\begin{figure*}[!t]
\begin{center}
\vspace{-1em}
\includegraphics[width=.8\linewidth]{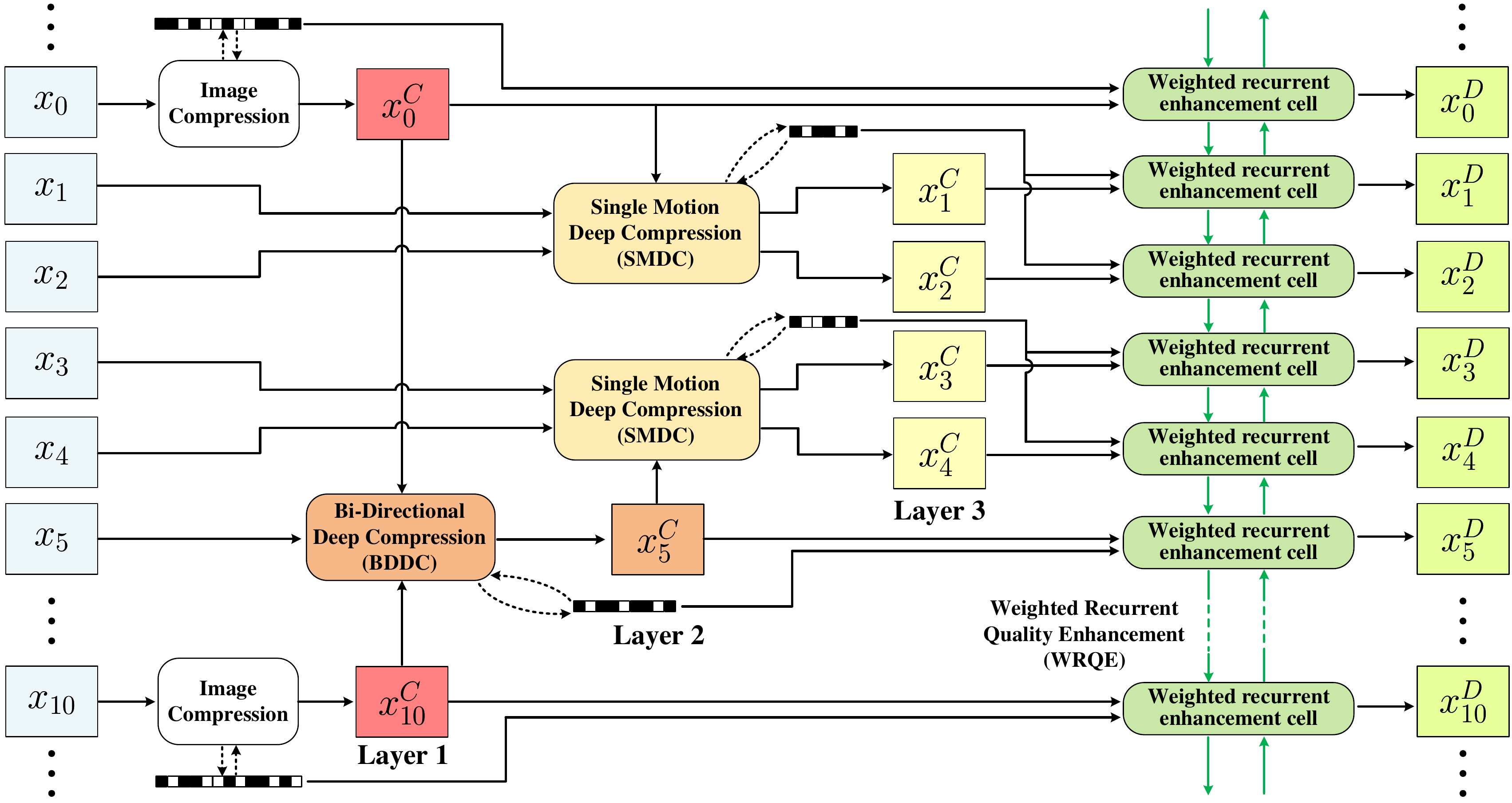}
\end{center}
\vspace{-1em}
\caption{The overall framework of our HLVC approach, which compresses video with three hierarchical quality layers using the proposed BDDC and SMDC networks, and employs the recurrent enhancement network WRQE in the deep decoder.
}\label{fig:framework}
\vspace{-1em}
\end{figure*}

\textbf{Deep video compression.} Based on the traditional image compression standards, several handcrafted algorithms, \eg, MPEG \cite{le1992mpeg}, H.264 \cite{wiegand2003overview} and H.265 \cite{sullivan2012overview}, were standardized for video compression.
In recent years, deep learning also attracted more attention in video compression. Many approaches~\cite{xu2018reducing, liu2018one, dai2017convolutional, li2019densenet, li2019deep} were proposed to replace the components in traditional video codecs by DNNs. For instance,
Liu~\etal~\cite{liu2018one} utilized a DNN in the fractional interpolation of motion compensation, and \cite{dai2017convolutional, li2019densenet, li2019deep} use DNNs to improve the in-loop filter. However, these methods only advance the performance of one particular module, and each module in video compression framework cannot be jointly optimized.

Most recently, several end-to-end deep video compression methods have been proposed~\cite{chen2019learning,chen2017deepcoder,wu2018video,cheng2019learning, lu2019dvc, habibian2019video}. Specifically,
Wu~\etal~\cite{wu2018video} proposed predicting frames by interpolation from reference frames, and the image compression network of~\cite{toderici2017full} is applied to compress the residual. In 2019, Lu~\etal~\cite{lu2019dvc} proposed the Deep Video Compression (DVC) method, in which optical flow is used to estimate the temporal motion, and two auto-encoders are employed to compress the motion and residual, respectively. Meanwhile, in~\cite{cheng2019learning}, spatial-temporal energy compaction is added into the loss function to improve the performance of video compression. Later, Habibian~\etal~\cite{habibian2019video} proposed the rate-distortion auto-encoder, which uses an autoregressive prior for video entropy coding.

Among the existing methods, only Wu~\etal~\cite{wu2018video} uses hierarchical prediction. Nevertheless, none of them learns to compress video with hierarchical quality, and therefore they fail to provide high quality references for the compression of other frames, and cannot take advantage of high quality information in multi-frame post-processing.

\textbf{Enhancement of compressed video.} Since lossy video compression inevitably leads to artifacts and quality loss, some works focus on enhancing the quality of compressed video \cite{wang2017novel, yang2017decoder, yang2018enhancing, yang2018multi, yang2019quality, wang2018multi, lu2018deep}. Among them, \cite{wang2017novel, yang2017decoder, yang2018enhancing} are single frame approaches with the input of one frame each time. Then, Yang~\etal~\cite{yang2018multi, yang2019quality} proposed multi-frame quality enhancement approaches, which make use of the inter-frame correlation.
Besides, the deep Kalman filter was proposed in \cite{lu2018deep} to reduce compression artifacts.

Nevertheless, above methods are all designed as post-processing modules for traditional video coding standards. Therefore, in the multi-frame approaches \cite{yang2018multi, yang2019quality}, the accurate frame quality cannot be obtained, and only can be estimated with prediction error.
In our HLVC approach, the compression quality of each frame is encoded into the bit stream, which is input together with the compressed frames to our enhancement network, making enhancement guided by accurate frame quality and as a component of our deep decoder in the whole video compression framework.

\section{The proposed approach}
\subsection{Framework} \label{framework}

Figure~\ref{fig:framework} shows the framework of our HLVC approach on the first Group Of Picture (GOP), and our framework is the same for each GOP. In HLVC, the frames are compressed as three hierarchical quality layers, namely layers 1, 2 and 3, with decreasing quality.

\textbf{Layer 1. }
The first layer (red frames in Figure~\ref{fig:framework}) is encoded by image compression method\footnote{For compressing the frames in layer 1, we use BPG~\cite{BPG} and Lee~\etal~\cite{lee2019context} in our PSNR and MS-SSIM models, respectively.}, and $x_i^C$ denotes the compressed frames. Similar to the ``I-frames'' in traditional codecs~\cite{wiegand2003overview,sullivan2012overview}, the frames in layer 1 consume the highest bit-rates, and are of the highest compression quality. As such, they are able to stop the error propagation during video encoding and decoding. More importantly, these frames provide high quality information, which benefits the compression and enhancement of neighboring frames.

\begin{figure*}[!t]
\begin{center}
\vspace{-1em}
\includegraphics[width=.9\linewidth]{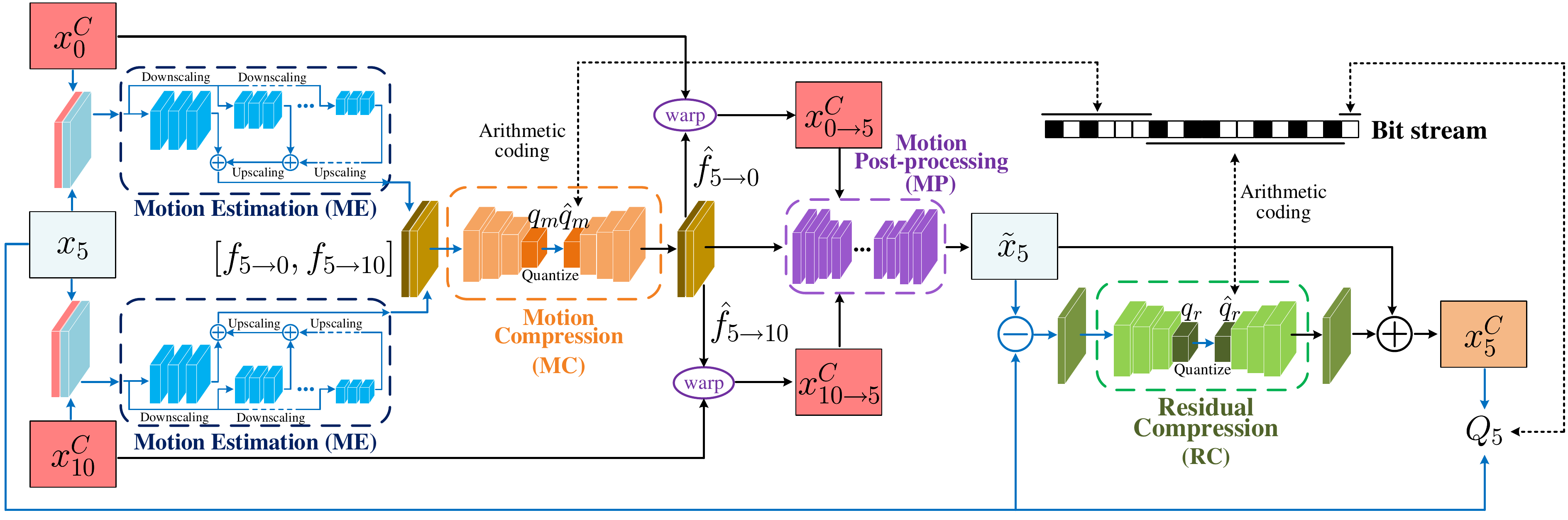}
\end{center}
\vspace{-.5em}
\caption{The architecture of our BDDC network. The blue arrows
indicate the procedures not included in the decoder.}\label{fig:bi_dirct}
\vspace{-1em}
\end{figure*}

\textbf{Layer 2. }
Then, the frames of layer 2 (orange frames in Figure~\ref{fig:framework}) are in the middle of two frames of layer 1. We propose the Bi-directional Deep Compression (BDDC) network to compress layer 2. Our BDDC network takes the previous and the upcoming compressed frames from layer 1 as bi-directional references.
We compress layer 2 as the medium quality layer, which also provides beneficial information to compress and enhance the low quality frames in layer 3. The BDDC network is introduced in Section~\ref{bi}.

\textbf{Layer 3. }
The remaining frames belong to layer 3 (yellow frames in Figure~\ref{fig:framework}), which are compressed with the lowest quality and contribute the least bit-rate. In the latest deep video compression approaches, \eg, Wu~\etal~\cite{wu2018video} and DVC~\cite{lu2018deep}, each frame requires at least one motion map for motion compensation. However, the motions between continuous frames are correlated, thus encoding one motion map for each frame leads to redundancy. Hence, we propose the Single Motion Deep Compression (SMDC) network, which applies a single motion map to describe the motions between multiple frames, and therefore the bit-rate can be reduced. Note that the frames $x_6$ to $x_9$ are compressed in the same manner as $x_1$ to $x_4$, so they are omitted in Figure~\ref{fig:framework}. The SMDC network is introduced in Section~\ref{sf}.

\textbf{Enhancement.}
Then, because of the high correlation among video frames~\cite{yang2018multi}, we develop the Weighted Recurrent Quality Enhancement (WRQE) network, in which the recurrent cells are weighted by quality features to reasonably leverage multi-frame information. Particularly, The quality of layer 3 can be significantly improved by taking advantage of the high quality information in layers 1 and 2. Since no additional information needs to be stored for improving the quality, this is equivalent to saving bit-rate, especially on low quality frames. Note that WRQE is a part of our deep decoder, with the inputs of both the compressed frames and the quality information encoded in the bit stream.
The WRQE network is detailed in Section~\ref{enh}.

\subsection{Bi-Directional Deep Compression (BDDC)} \label{bi}

The BDDC network for compressing layer 2 is shown in Figure~\ref{fig:bi_dirct}. Here, we also use the first GOP as an example.
In BDDC, we first use the \textbf{Motion Estimation (ME)} subnet to capture the temporal motion between the reference and target frames. Since the interval between the frames in layers 1 and 2 is long (\eg, 5 frames in Figure~\ref{fig:bi_dirct}), we follow \cite{ranjan2017optical} to apply a pyramid network to handle the large motions, taking advantage of the large receptive field.
Note that we use backward warping in our approach, and therefore we estimate backward motions. For example, in Figure~\ref{fig:bi_dirct}, the outputs of our ME subnet are the motions from $x_5$ to $x^C_0$ (denoted as $f_{5\rightarrow 0}$) and from $x_5$ to $x^C_{10}$ (denoted as $f_{5\rightarrow 10}$), respectively.

Given the estimated motions, an auto-encoder is utilized for \textbf{Motion Compression (MC)}.
Because of the similarity among video frames, there exists correlation between the motions of different frames. Therefore, we propose concatenating (denoted as $[\cdot, \cdot, ...]$) the bi-directional motions as the input to the encoder $E_m$, which transforms the input to a latent representation $q_m$. Then, $q_m$ is quantized to $\hat{q}_m$, and $\hat{q}_m$ is fed to the decoder $D_m$ to generate compressed motion. Here, $\hat{q}_m$ is encoded to bits by arithmetic coding~\cite{langdon1984introduction}. As such, defining $\hat{f}_{5\rightarrow 0}$ and $\hat{f}_{5\rightarrow 10}$ as the compressed motions, our MC subnet can be formulated as
\begin{gather}
    \hat{q}_m = \text{round}(E_m([f_{5\rightarrow 0}, f_{5\rightarrow 10}])), \label{mc} \\
    [\hat{f}_{5\rightarrow 0}, \hat{f}_{5\rightarrow 10}] = D_m(\hat{q}_m). \label{md}
\end{gather}

Next, the reference frames $x_0^C, x_{10}^C$ are warped to the target frame using the compressed motions. The \textbf{Motion Post-processing (MP)} subnet then merges the warped frames for motion compensation. Defining $W_b$ as the backward warping operation, the motion compensation can be formulated as
\begin{gather}
    x^C_{0\rightarrow 5} = W_b(x^C_0, \hat{f}_{5\rightarrow 0}), \ \
    x^C_{10\rightarrow 5} = W_b(x^C_{10}, \hat{f}_{5\rightarrow 10}), \label{warp}\\
    \tilde{x}_5 = M\hspace{-1.7pt}P([x^C_{0\rightarrow 5}, x^C_{10\rightarrow 5}, \hat{f}_{5\rightarrow 0}, \hat{f}_{5\rightarrow 10}]), \label{mp}
\end{gather}
where $\tilde{x}_5$ denotes the compensated frame.
Finally, the residual between the compensated frame $\tilde{x}_5$ and the raw frame $x_5$ is compressed by the \textbf{Residual Compression (RC)} subnet. Similar to the MC subnet, there are encoder ($E_r$) and decoder ($D_r$) networks in RC. Using $\hat{q}_r$ to denote the quantized latent representation, the RC subnet can be written as
\begin{gather}
    \hat{q}_r = \text{round}(E_r(x_5 - \tilde{x}_5)), \label{rc} \\
    x_5^C = D_r(\hat{q}_r) + \tilde{x}_5, \label{rc2}
\end{gather}
where $x_5^C$ represents the compressed frame of $x_5$. In RC, $\hat{q}_r$ is encoded to bits using arithmetic coding, which contributes to the bits of layer 2 together with the encoded $\hat{q}_m$. Besides, the compression quality $Q_5$ is calculated and included in the bit stream, and is to be used in our deep decoder (in Section~\ref{enh}). In this paper, Multi-Scale Structural SIMilarity (MS-SSIM)~\cite{wang2003multiscale} and Peak Signal-to-Noise Ratio (PSNR) are used to evaluate quality. The details of each subnet are shown in the \textit{Supplementary Material}\footnote{\url{https://arxiv.org/abs/2003.01966}.}.

\begin{figure}[!t]
\begin{center}
\includegraphics[width=.95\linewidth]{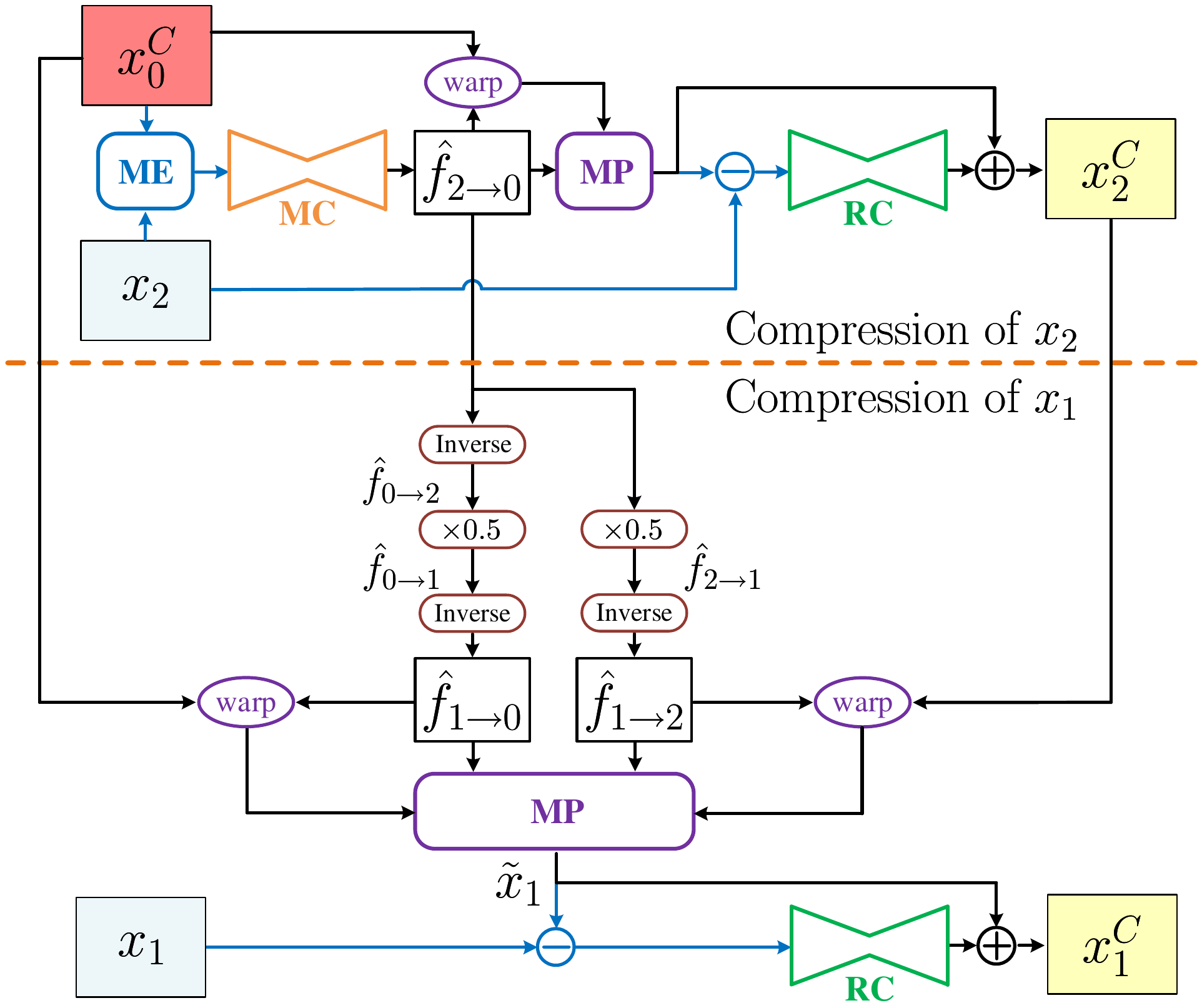}
\end{center}
\vspace{-.5em}
\caption{The architecture of our SMDC network.
The bit stream is generated from MC and RC, and also includes the compression quality. This figure omits the bit stream for simplicity.}\label{fig:flow}
\vspace{-1em}
\end{figure}

\subsection{Single Motion Deep Compression (SMDC)} \label{sf}

In the following, the remaining frames are compressed as layer 3 by the proposed SMDC network, using the nearest compressed frames in layers 1 and 2 as references. In our SMDC network, we compress two frames with a single motion map. For instance, as illustrated in Figure~\ref{fig:framework}, the frames $x_1$ and $x_2$ are compressed using a single motion map with the reference of $x_0^C$, and $x_3$ and $x_4$ use $x_5^C$ as reference.

\begin{figure*}[!t]
\begin{center}
\vspace{-1em}
\includegraphics[width=.99\linewidth]{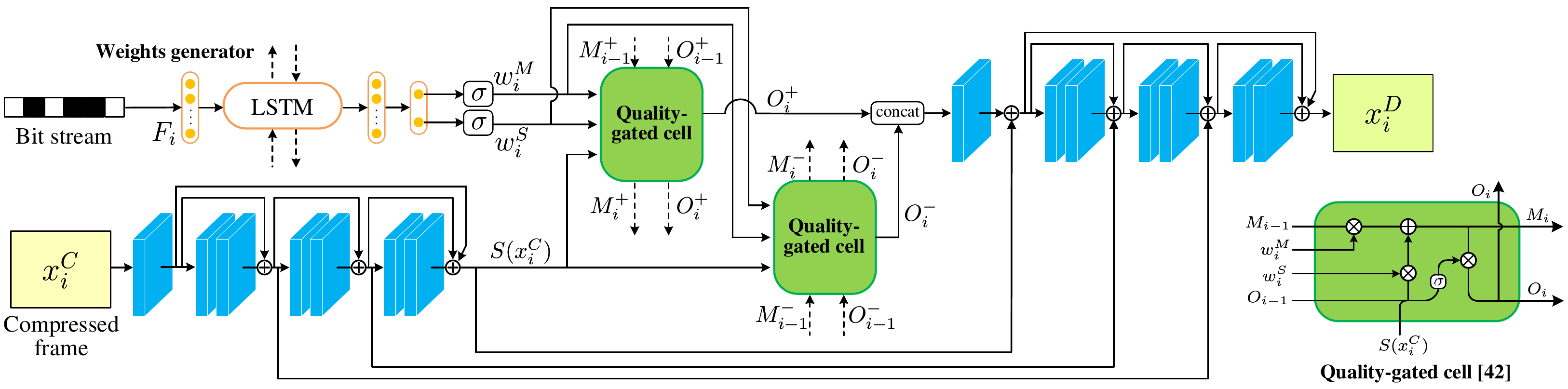}
\end{center}
\vspace{-.5em}
\caption{The architecture of our WRQE network. The black dash lines indicate the information from previous cells or to subsequent cells.}\label{fig:enh}
\vspace{-1em}
\end{figure*}

Figure~\ref{fig:flow} shows the architecture of our SMDC network on $x_1$ and $x_2$ as an example.
We can see from Figure~\ref{fig:flow} that the frame $x_2$ is first compressed by a DNN with similar architecture as BDDC, which contains the ME, MC, MP and RC subnets, and the compressed frame $x_2^C$ is obtained.
As mentioned above, due to the correlation of motions among multiple neighboring frames, we propose using the motion between $x_0^C$ and $x_2$ to predict the motions between $x_1$ and $x_0^C$ or $x_2^C$. As such, the frame $x_1$ can be compressed with the reference frames of $x_0^C$ and $x_2^C$, without bits consumption for motion map, thus improving the rate-distortion performance.

In our SMDC network, we propose applying the \textbf{inverse motion} for motion prediction. Specifically, a motion map can be defined as $f(a,b)=[\Delta a(a,b), \Delta b(a, b)]$, where $a$ and $b$ denote the coordinates, while $\Delta a$ and $\Delta b$ are the horizontal and vertical motion maps, respectively. For $f(a,b)$, the inverse motion can be expressed as
\begin{equation}
    f_{\text{inv}}\big(a+\Delta a(a,b),b+\Delta b(a, b)\big) = -f(a,b). \label{inv}
\end{equation}
In \eqref{inv}, $f(a,b)$ describes that the pixel at $(a, b)$ moves to the new position of $\big(a+\Delta a(a,b),b+\Delta b(a, b)\big)$, and therefore the value of $-f(a,b)$ should be assigned to $f_{\text{inv}}$ at the new position. For simplicity, we define the inverse operation as $\text{Inverse}(\cdot)$, \ie, $f_{\text{inv}} = \text{Inverse}(f)$.

Recall that \textbf{backward warping} is adopted in our approach, so the motion for compressing $x_2$ is from $x_2$ to $x_0^C$, which is defined as $f_{2\rightarrow 0}$. Similarly, using $x_0^C$ and $x_2^C$ as reference frames, the motions from $x_1$ to $x_0^C$ (denoted as $\hat{f}_{1\rightarrow 0}$) and from $x_1$ to $x_2^C$ (denoted as $\hat{f}_{1\rightarrow 2}$) are required for the compression of $x_1$. Note that, since the raw frame $x_2$ is not available at the decode side, $f_{2\rightarrow 0}$ cannot be recovered in decoding. Hence, the compressed motion $\hat{f}_{2\rightarrow 0}$ is used to predict $\hat{f}_{1\rightarrow 0}$ and $\hat{f}_{1\rightarrow 2}$. Given $\hat{f}_{2\rightarrow 0}$ and the inverse operation, $\hat{f}_{1\rightarrow 0}$ can be predicted as
\begin{equation}\label{inv1}
    \hat{f}_{1\rightarrow 0} =
    \text{Inverse}(
    \underbrace{0.5\times
    \underbrace{\text{Inverse}(\hat{f}_{2\rightarrow 0})}_{\let\scriptstyle\textstyle\substack{\hat{f}_{0\rightarrow 2}}}}
    _{\let\scriptstyle\textstyle\substack{\hat{f}_{0\rightarrow 1}}}
    ).
\end{equation}
In a similar way, $\hat{f}_{1\rightarrow 2}$ is obtained by
\begin{equation}\label{inv2}
    \hat{f}_{1\rightarrow 2} =
    \text{Inverse}(
    \underbrace{0.5\times
    { \hat{f}_{2\rightarrow 0}}}
    _{\let\scriptstyle\textstyle\substack{\hat{f}_{2\rightarrow 1}}}
    ).
\end{equation}
Then, the same as \eqref{warp} and \eqref{mp}, the reference frames are warped and fed into the MP subnet together with the predicted motions to generate the motion compensated frame $\tilde{x}_1$. Finally, the residual ($x_1-\tilde{x}_1$) is compressed by the RC subnet to obtain the compressed frame $x^C_1$. Here, the compressed quality $Q_1$ and $Q_2$ are also included in the bit stream, which are to be utilized in our WRQE network.

\subsection{Weighted Recurrent Enhancement (WRQE)} \label{enh}

Finally, at the decoder side, we use WRQE for quality enhancement. The WRQE network is designed based on the QG-ConvLSTM method~\cite{yang2019quality} with a spatial-temporal structure, which uses a quality-gated cell to exploit multi-frame correlations. The architecture is illustrated in Figure~\ref{fig:enh}.

Different from~\cite{yang2019quality}, we adopt residual blocks in the spatial feature extraction and reconstruction networks, and employ skip connections to improve the enhancement performance. More importantly, as discussed in~\cite{yang2018multi, yang2019quality}, the significance of a frame for enhancing other frames depends on its relative quality compared to others. However, in~\cite{yang2018multi, yang2019quality}, the accurate quality of each frame cannot be obtained in the decoder.
In contrast, since the compression quality is encoded in our bit stream, we have access to the compression quality $Q_i$ and bit-rate $B_i$ from our bit stream,
and we utilize $F_i = \{Q_{j}, B_{j}\}_{j=i-2}^{i+2}$ as the ``quality feature''. We feed $F_i$ into the weights generator, and obtain the weights $w_i = [w^M_i, w^S_i]$, which are input to the quality-gated cell~\cite{yang2019quality} together with the spatial features $S(x^C_i)$.

\begin{figure*}[!t]
\centering
\vspace{-1em}
\subfigure{\includegraphics[width=.265\linewidth]{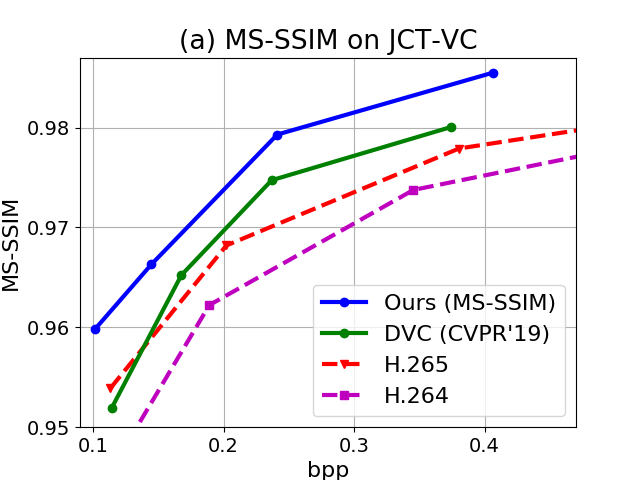}}
\hspace{-1.5em}
\subfigure{\includegraphics[width=.265\linewidth]{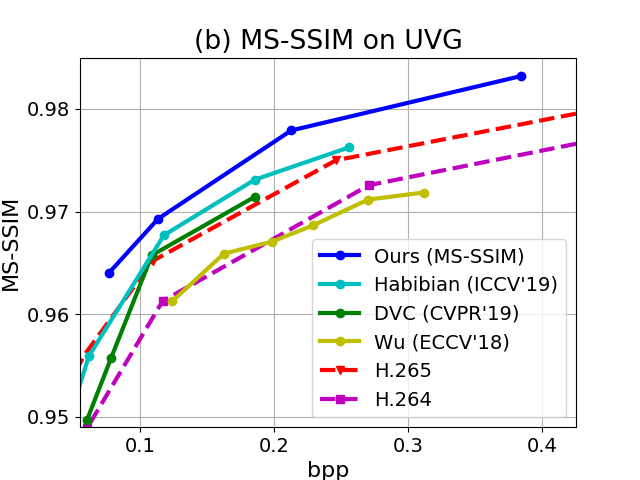}}
\hspace{-1.5em}
\subfigure{\includegraphics[width=.265\linewidth]{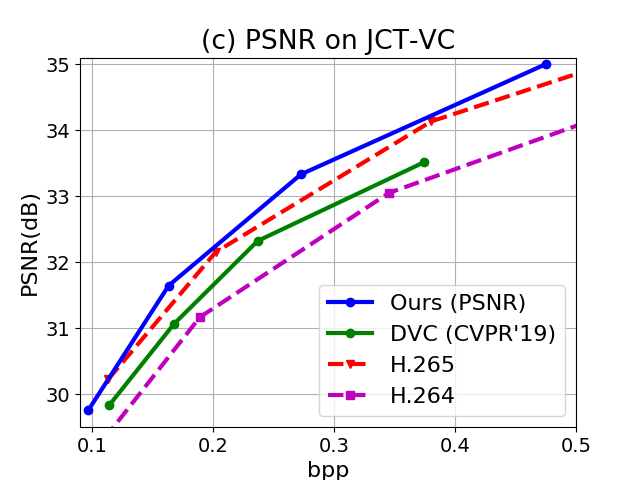}}
\hspace{-1.5em}
\subfigure{\includegraphics[width=.265\linewidth]{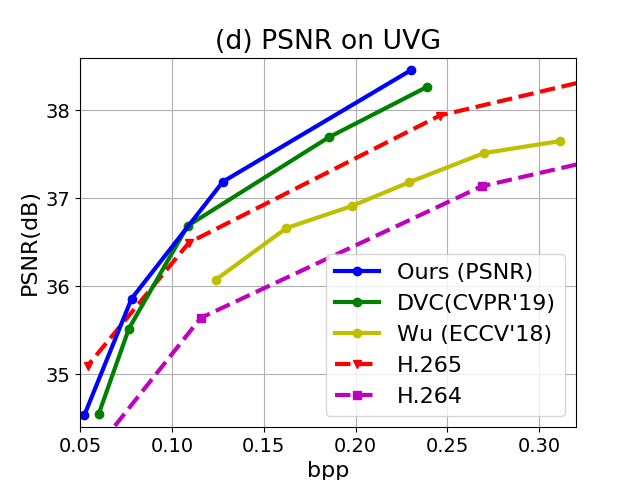}}
\caption{Rate-distortion performance in terms of MS-SSIM and PSNR.}
\vspace{-1em}
\label{fig:results}
\end{figure*}

As shown in Figure \ref{fig:enh}, the weights $w_i = [w^M_i, w^S_i]$ are learned to reasonably control $M_i$ to forget previous memory and update the current information.
Specifically, on high quality frames, the memory $M_i$ is expected to be multiplied with a small $w^M_i$ to forget previous low quality information, but the update weight $w^S_i$ is expected to be large, to add its high quality information to the memory for enhancing other frames.
In contrast, a large $w^M_i$ and a small $w^S_i$ are expected on low quality frames.
Furthermore, since $w_i^M$ is the output of a sigmoid function, $w_i^M<1$ holds, and thus the
information from previous frames decreases in the memory as frame distance increases.
This matches the fact the frames with longer distance are less correlated, and therefore are less useful for quality enhancement. As such, in the quality-gated cell, the frames with different quality contribute to the memory $M_i$ with different significance, making our WRQE network reasonably leverage multi-frame information for quality enhancement.

\subsection{Training strategy} \label{train}

In the training phase, we use the density model of \cite{balle2017end} to estimate the bit-rate for encoding $\hat{q}_m$ and $\hat{q}_r$ in \eqref{mc} and \eqref{rc}, and define the estimated bit-rate as $R(\cdot)$. Then, we follow \cite{mentzer2018conditional, lu2019dvc} to formulate the loss as
\begin{equation}
    L = \lambda D + R, \label{RDO}
\end{equation}
in which $\lambda$ is the hyperparameter to control the trade-off between distortion $D$ and bit-rate $R$.

It can be seen from \eqref{RDO} that the compression quality of the trained model depends on the hyperparameter $\lambda$, \ie, larger $\lambda$ results in higher quality and higher bit-rate. Therefore, to achieve the hierarchical compression quality in our HLVC approach, different $\lambda$ values are applied for our BDDC and SMDC networks, which compress layers 2 and 3, respectively.
To be specific, given \eqref{RDO} and the estimated bit-rates, we set the loss function of our BDDC network as
\begin{equation}
    L_\text{BD} = \lambda_\text{BD}\cdot
    \underbrace{D(x_5, x_5^C)}_\text{Distortion}  +
    \underbrace{R(\hat{q}_m) + R(\hat{q}_r)}_\text{Total bit-rate},
    \label{l1}
\end{equation}
and the loss for our SMDC network is
\begin{multline}
    L_\text{SM} = \lambda_\text{SM}\cdot
    \underbrace{\big(D(x_1, x_1^C) + D(x_2, x_2^C)\big)}_\text{Total distortion} \\
    + \underbrace{R(\hat{q}_m) + R(\hat{q}_{r1}) + R(\hat{q}_{r2})}_\text{Total bit-rate}.
    \label{l2}
\end{multline}
In \eqref{l2}, $\hat{q}_{r1}$ and $\hat{q}_{r2}$ are the representations in the RC networks of $x_1$ and $x_2$, respectively.
In \eqref{l1} and \eqref{l2}, we use the Mean Square Error (MSE) as the distortion, \ie, $D(x,y) = \text{MSE}(x,y)$, when training our HLVC approach for PSNR. We apply $D(x, y) = 1-\text{MS-SSIM}(x,y)$ when optimizing for MS-SSIM.
More importantly, we set $\lambda_\text{BD} > \lambda_\text{SM}$ in \eqref{l1} and \eqref{l2} to make our approach learn to compress layer 2 with higher quality than layer 3, thus achieving the hierarchical quality.

Finally, we train our WRQE network by minimizing the loss function of
\begin{equation}
    L_\text{QE} = \frac{1}{N}\sum_{i=1}^N D(x_i, x_i^D), \label{l3}
\end{equation}
where $N$ is the step length of our recurrent network. Because of the bi-directional recurrent structure, larger $N$ leads to longer decoding latency and also longer training time. Therefore, we set $N$ as 11 (the interval of frames in layer 1) in both training and inference phases.

\begin{table*}[!t]
\vspace{-2em}
\centering
\begin{threeparttable}
\footnotesize
\caption{BDBR with the anchor of H.265 (x265 ``LDP very fast''). Bold indicates best results.}\label{tab:bdbr}
\begin{tabular}{|c|rrr|rrrr|rrrr|}
\hline
\multirow{4}[1]{*}{Dataset} & \multicolumn{7}{c|}{BDBR (\%) calculated by MS-SSIM} & \multicolumn{4}{c|}{BDBR (\%) calculated by PSNR}\\
\cline{2-12}
 & \multicolumn{3}{c|}{Optimized for PSNR} &\multicolumn{4}{c|}{Optimized for MS-SSIM} & \multicolumn{4}{c|}{Optimized for PSNR}\\
\cline{2-12}
&  \multicolumn{1}{c}{Wu $\dagger$}  & \multicolumn{1}{c}{DVC} & \multicolumn{1}{c|}{HLVC} &   \multicolumn{1}{c}{Cheng $\ddagger$} & \multicolumn{1}{c}{Habibian} &\multicolumn{1}{c}{HLVC} &
\multicolumn{1}{c|}{HLVC}& \multicolumn{1}{c}{Wu $\dagger$} & \multicolumn{1}{c}{\ \ DVC \ \ }   & \multicolumn{1}{c}{HLVC} &\multicolumn{1}{c|}{HLVC}\\
& \multicolumn{1}{c}{\cite{wu2018video}} & \multicolumn{1}{c}{\cite{lu2019dvc}}& \multicolumn{1}{c|}{(Ours)} & \multicolumn{1}{c}{\cite{cheng2019learning}}  &  \multicolumn{1}{c}{\cite{habibian2019video}}& w/o WRQE&
\multicolumn{1}{c|}{(Ours)}&\multicolumn{1}{c}{\cite{wu2018video}} & \multicolumn{1}{c}{\cite{lu2019dvc}} &w/o WRQE &
\multicolumn{1}{c|}{(Ours)}
 \\
\hline

UVG &  $49.42$ & $\mathbf{8.05}$ & $11.24$ &  - & $3.71$& $-21.94$ &{$\mathbf{-30.12}$} & $40.29$ & $8.89$ & $9.39$&$\mathbf{-1.37}$\\

Class B &  -&$-2.74$ & $\mathbf{-8.11}$&-&-& $-35.32$ &{$\mathbf{-37.44}$}&- & $1.98$ & $-2.61$& {$\mathbf{-11.75}$}\\

Class C &  -&{$-6.88$}&{$\mathbf{-9.10}$} &$3.48$ &-& $-20.87$ & {$\mathbf{-23.63}$}  & - & $25.88$ &$20.44$ &$\mathbf{7.83}$ \\

Class D &  -&{$\mathbf{-18.51}$}&{$-18.44$}&{$-23.72$}&-& $-32.94$ &{$\mathbf{-52.56}$} &- & $15.34$ & $-1.52$ &{$\mathbf{-12.57}$}\\

\hline
Average &  - & $-5.02$ &$\mathbf{-6.10}$& - & - & $-27.77$& {$\mathbf{-35.94}$}  &-&$13.03$ & $6.43$&{$\mathbf{-4.46}$}\\

\hline

\end{tabular}%
\begin{tablenotes}
\item $\dagger$ Wu \etal \cite{wu2018video} does not provide the result on each video, and therefore the BDBR values of \cite{wu2018video} are calculated by the average curves in Figure \ref{fig:results}.
\item $\ddagger$ The results of Cheng \etal \cite{cheng2019learning} are calculated by the data provided by the authors, which are tested on the first 81 frames of each video.
\end{tablenotes}
\end{threeparttable}

\vspace{-1.5em}
\end{table*}%

\section{Experiments}
\subsection{Settings}

Our BDDC and SMDC networks are trained on the Vimeo-90k~\cite{xue2019video} dataset, and we collect 142 videos from Xiph~\cite{Xiph} and VQEG~\cite{VQEG} to train our WRQE network. We test HLVC on the JCT-VC~\cite{bossen2013common} (Classes B, C and D) and the UVG~\cite{UVG} datasets, which are not overlapping with our training sets. Among them, the UVG and JCT-VC Class B are high resolution ($1920\times 1080$) datasets\footnote{Since our entropy model requires each dimension to be a multiple of 16, we crop the height to 1072 by cutting the bottom 8 pixels.}, and the JCT-VC Classes C and D have resolutions of $832\times 480$ and $416\times 240$, respectively. For a fair comparison with \cite{lu2019dvc}, we follow \cite{lu2019dvc} to test JCT-VC videos on the first 100 frames, and test UVG videos on all frames. The quality is evaluated in terms of MS-SSIM and PSNR. We train the models with $\lambda_\text{SM} = 8, 16, 32, 64$ for MS-SSIM, and with $\lambda_\text{SM} = 256, 512, 1024, 2048$ for PSNR. To achieve hierarchical quality, we set $\lambda_\text{BD} = 4\times \lambda_\text{SM}$.
We compare HLVC with the latest learned video compression methods. Among them, Habibian~\etal~\cite{habibian2019video}  and Cheng~\etal~\cite{cheng2019learning}  are optimized for MS-SSIM. DVC~\cite{lu2019dvc}  and Wu~\etal~\cite{wu2018video}  are optimized for PSNR. Furthermore, we include the video coding standards H.264~\cite{wiegand2003overview} and H.265~\cite{sullivan2012overview} in our comparisons. We follow \cite{lu2019dvc} to use x264 and x265 ``LDP very fast'' mode, with the same the GOP size as our approach (\ie, GOP = 10) on all videos. The results are presented in this section.

Please refer to the \textit{Supplementary Material}\footnote{\url{https://arxiv.org/abs/2003.01966}.} for more experimental results, including visual results, different GOP sizes, the comparison with other configurations of x265, \etc.

\subsection{Results}

\textbf{Rate-distortion curve.} Figure~\ref{fig:results} demonstrates the rate-distortion curves on the JCT-VC and UVG datasets. The quality is evaluated in terms of MS-SSIM and PSNR, and the bit-rate is calculated by bits per pixel (bpp). As shown in Figure~\ref{fig:results} (a) and (b), our MS-SSIM model outperforms all learned approaches, and reaches better performance than H.264 and H.265. Especially, at low bit-rate on UVG, Habibian~\etal~\cite{habibian2019video} is comparable with H.265 and DVC~\cite{lu2019dvc} performs worse than H.265. On JCT-VC, DVC~\cite{lu2019dvc} is only comparable with H.265 at low bit-rate. On the contrary, the rate-distortion curves of our HLVC approach are obviously above H.265 from low to high bit-rates. The PSNR curves are illustrated in Figure~\ref{fig:results} (c) and (d). It can be seen that our PSNR model achieves better performance than the latest PSNR optimized methods DVC~\cite{lu2019dvc} and Wu~\etal~\cite{wu2018video}, and also outperforms H.265 on the JCT-VC dataset. On UVG, we reach better performance than H.265 at high bit-rate.

\textbf{Bit-rate reduction.} Furthermore, we evaluate the Bj{\o}ntegaard Delta Bit-Rate (BDBR)~\cite{bjontegaard} with the anchor of H.265. BDBR calculates the average bit-rate difference in comparison with the anchor, and lower BDBR values indicate better performance. Table~\ref{tab:bdbr} shows BDBR calculated by MS-SSIM and PSNR, in which the negative numbers indicate reducing bit-rate compared to the anchor, thus outperforming H.265, and the bold numbers are the best results among all learned methods.

In Table \ref{tab:bdbr}, for a fair comparison on MS-SSIM with the PSNR optimized methods DVC~\cite{lu2019dvc} and H.265, we first report the BDBR of our PSNR model in terms of MS-SSIM. As Table~\ref{tab:bdbr} shows, our PSNR model outperforms H.265 on MS-SSIM with the average BDBR of $-6.10\%$, which is also better than DVC (BDBR = $-5.02\%$). On JCT-VC Class C, our PSNR model even obviously outperforms the MS-SSIM optimized method Cheng~\etal~\cite{cheng2019learning} in terms of MS-SSIM. Furthermore, our MS-SSIM model successfully outperforms all existing learned methods on MS-SSIM, and reduces the bit-rate of H.265 by $35.94\%$ on average. More importantly, the performance of our MS-SSIM model before quality enhancement (without WRQE) (BDBR = $-27.77\%$) is still significantly better than all previous methods. In conclusion, our HLVC approach achieves the state-of-the-art MS-SSIM performance among learned and handcrafted video compression methods.

Table~\ref{tab:bdbr} also shows the BDBR results calculated by PSNR. As shown in Table~\ref{tab:bdbr}, our PSNR model performs best among all learned methods in terms of PSNR. Especially, we outperform the latest PSNR method DVC~\cite{lu2019dvc} on all test sets. In comparison with H.265, our PSNR model reduces the bit-rate by $4.46\%$ on average, although having $7.83\%$ bit-rate overhead on JCT-VC Class C. Among the 20 videos in our test sets, our PSNR model beats H.265 on 14 videos in terms of PSNR. Besides, as shown in Table~\ref{tab:bdbr}, our PSNR model without WRQE still outperforms the latest PSNR method DVC~\cite{lu2019dvc}. In summary, our HLVC approach outperforms all existing learned methods on PSNR, and reaches better performance than H.265 (x265 ``LDP very fast'').

\begin{figure}[!t]
\begin{center}
\vspace{-.5em}
\includegraphics[width=.8\linewidth]{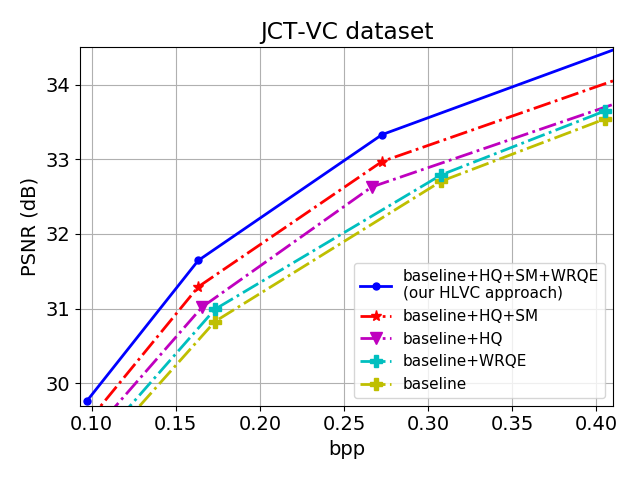}
\vspace{-1em}
\end{center}
\caption{Ablation studies of each component in our approach.}\label{fig:abl}
\vspace{-1.5em}
\end{figure}

\subsection{Ablation studies}

The ablation studies are conducted to prove the effectiveness of each component in our HLVC approach. We define the baseline model as our approach without Hierarchical Quality (HQ) (using the models trained with the same $\lambda$ for all frames), without the Single Motion (SM) strategy (compress one motion map for each frame) and without our enhancement network WRQE. Then, we analyze the performance of the baseline model and add these components successively, \ie, ``baseline+HQ'', ``baseline+HQ+SM'' and our whole framework ``baseline+HQ+SM+WRQE''. Moreover, we also discuss the enhancement on non-hierarchical video (``baseline+WRQE''). The ablation results are illustrated in Figure \ref{fig:abl}.

\begin{figure}[!t]
\begin{center}
\vspace{-1.5em}
\includegraphics[width=.9\linewidth]{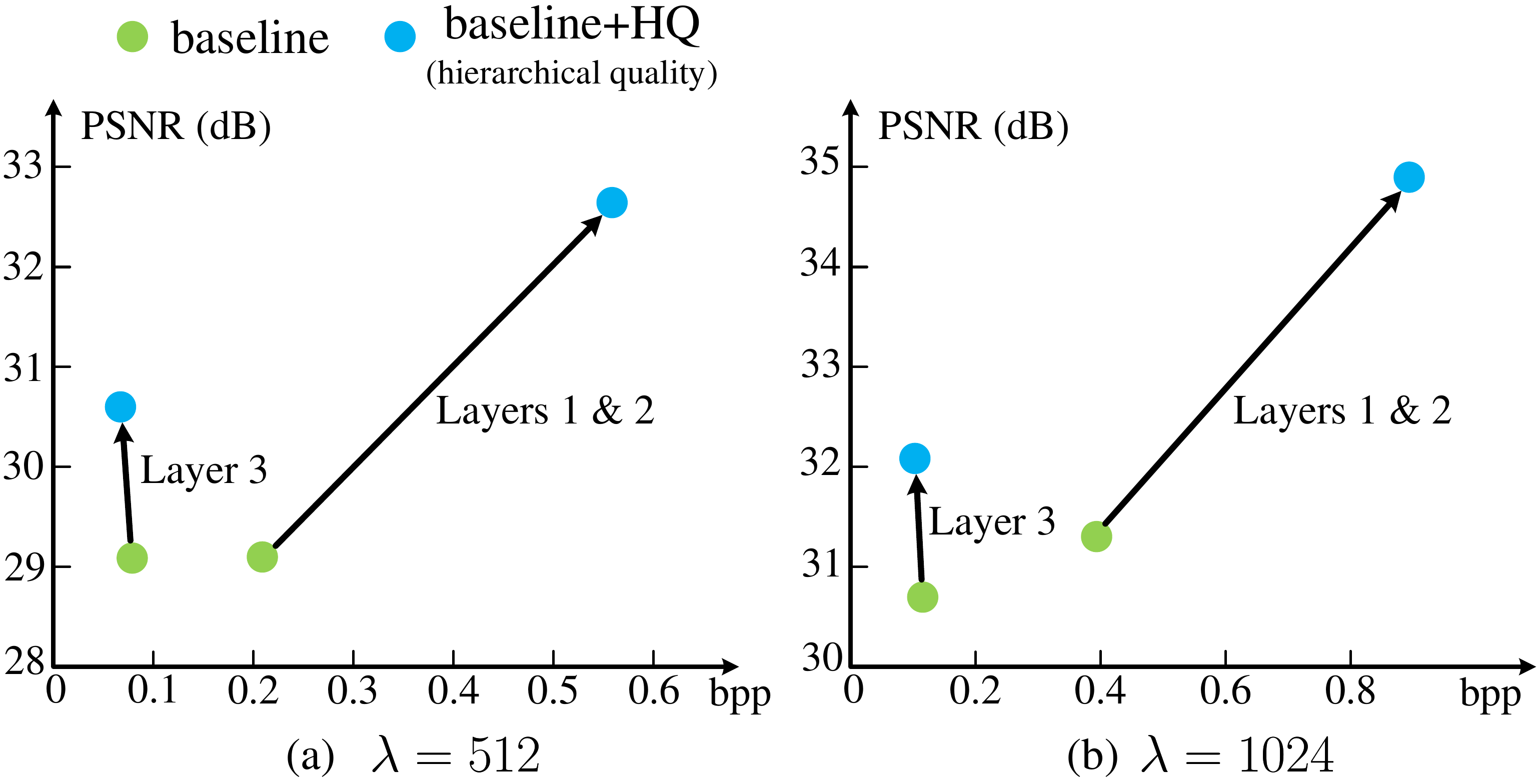}
\end{center}
\vspace{-1em}
\caption{Average bit-rate and PSNR on different layers.}\label{fig:hq}
\vspace{-1.5em}
\end{figure}

\textbf{Hierarchical quality.} Figure \ref{fig:abl} shows that ``baseline+HQ'' obviously outperforms the baseline model, indicating the effectiveness of applying the hierarchical quality to improve the compression performance. Besides, Figure \ref{fig:hq} shows the changes of bit-rate and PSNR on high quality (layers 1 and 2) and low quality (layer 3) frames from the baseline to ``baseline+HQ''. It can be seen that, on layers 1 and 2, employing hierarchical quality enlarges both the bit-rate and PSNR. On layer 3, ``baseline+HQ'' achieves higher PSNR than the baseline but even with lower bit-rate. This is because layers 1 and 2 in ``baseline+HQ'' provide a high quality reference for compressing layer 3. Since most frames of a video are in layer 3, applying hierarchical quality improves the compression performance.

\textbf{Single motion strategy.} Then, as shown in Figure~\ref{fig:abl}, adding SMDC (``baseline+HQ+SM'') further improves the performance compared to ``baseline+HQ'', by reducing the bits used for motion maps. For example, in ``baseline+HQ'', the average bit-rate for motion information is $0.0175$ bpp at $\lambda=256$, and the total bit-rate is $0.0973$ bpp. Using SMDC, the bits consumed for motion reduce to $0.0134$ bpp, which is $23.4\%$ lower than without SMDC, and the total bit-rate also decreases to $0.0969$ bpp. Meanwhile, PSNR improves from $29.26$ dB (``baseline+HQ'') to $29.47$ dB (``baseline+HQ+SM''), since more bits can be allocated on residual coding. This validates that our SMDC network successfully reduces the redundancy of video motion, and benefits the compression performance.

\begin{figure}[!t]
\begin{center}
\vspace{-1.5em}
\includegraphics[width=.9\linewidth]{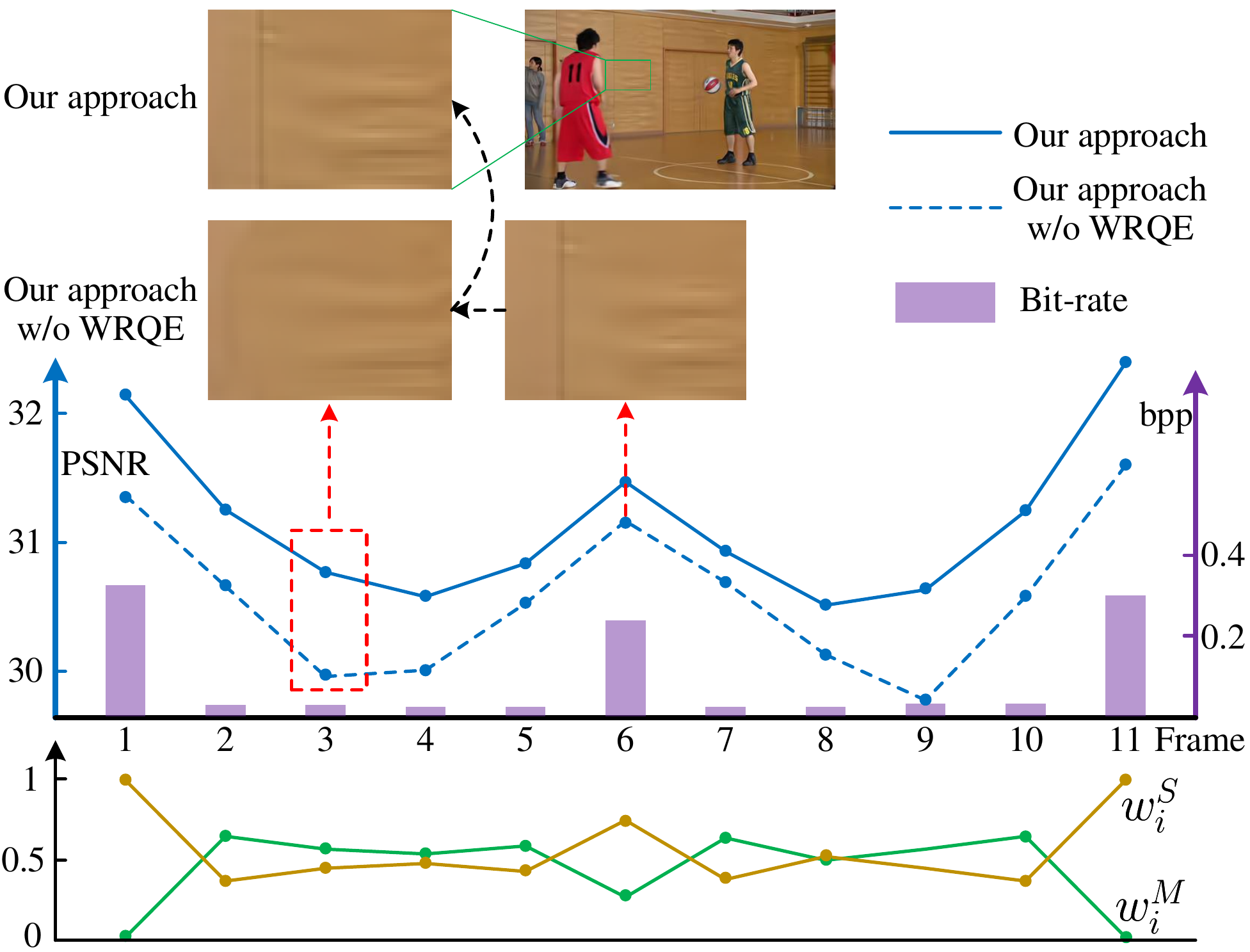}
\end{center}
\vspace{-1em}
\caption{Example results of WRQE on \textit{BasketballPass}.}\label{wrqe}
\vspace{-1em}
\end{figure}

\textbf{Recurrent enhancement.} As we can see from Figure~\ref{fig:abl}, our WRQE network (``baseline+HQ+SM+WRQE'') effectively further enhances the quality based on ``baseline+HQ+SM''.
As the example in Figure~\ref{wrqe} shows, our WRQE network significantly enhances compression quality, especially on low quality frames, \eg, the PSNR improvement is around 1 dB on frames 3 and 9. Figure~\ref{wrqe} also shows the learned weights of $w_i^S$ and $w_i^M$. It can be seen that on high quality frames, our WRQE network learns to generate larger $w_i^S$ and smaller $w_i^M$ to decrease the previous memory and update its helpful information to the memory, and the opposite for low quality frames. Moreover, as the visual results shown in Figure~\ref{wrqe} indicate, frame 3 surfers from severe distortion due to the low bit-rate, while frame 6 with higher quality are highly correlated with frame 3. Then, in WRQE, because of the large $w_i^S$ of frame 6, large proportion of its information is updated to the memory. Therefore, it is able to recover the lost information in frame 3 and significantly enhance the quality. These results validate the effectiveness of our WRQE network.

\textbf{Benefits of hierarchy for enhancement.} Finally, we show the result of our WRQE network on the baseline model without hierarchical quality. As shown in Figure~\ref{fig:abl}, the quality improvement from the baseline to ``baseline+WRQE'' is much less than that on our hierarchical quality method (``baseline+HQ+SM'' to our HLVC approach). This is caused by the similar quality on each frame in the non-hierarchical model, so there is no high quality reference to help the enhancement of other frames. This shows that the proposed hierarchical quality structure facilitates our WRQE network on enhancement, and as mentioned above, our WRQE network also successfully learns to reasonably make use of the hierarchical quality. As a result, our whole framework achieves the state-of-the-art performance among learned video compression methods, and outperforms H.265 (x265 ``LDP very fast'').

\vspace{-.3em}
\section{Conclusion and future work}
\vspace{-.1em}

This paper has proposed a learned video compression approach with hierarchical quality and recurrent enhancement. To be specific, we proposed compressing frames in the hierarchical layers 1, 2 and 3 with decreasing quality, using an image compression method for the first layer and the proposed BDDC and SMDC networks for the second and third layers, respectively. We developed the WRQE network with inputs of compressed frames, quality and bit-rate information, for multi-frame enhancement. Our experiments validated the effectiveness of our HLVC approach.

The same as other learned video compression methods, we manually set the frame structure in our approach.
A promising direction for future work is to develop DNNs which learn to automatically design the prediction and hierarchical structures.

\vspace{0.5em}

\noindent {\small \textbf{Acknowledgments.} This work was partly supported by ETH Zurich Fund (OK), Amazon through an AWS grant, and Nvidia through a GPU grant.}
{\small
\bibliographystyle{ieee_fullname}
\bibliography{egbib}
}

\newpage
\twocolumn[{%
\maketitle
\centering
\section*{Learning for Video Compression with Hierarchical Quality and Recurrent Enhancement \\--Supplementary Material--}
\vspace{1em}
}]

\section{Details of our framework}
\subsection{The BDDC network}

\begin{figure}[!h]
\begin{center}
\includegraphics[width=1\linewidth]{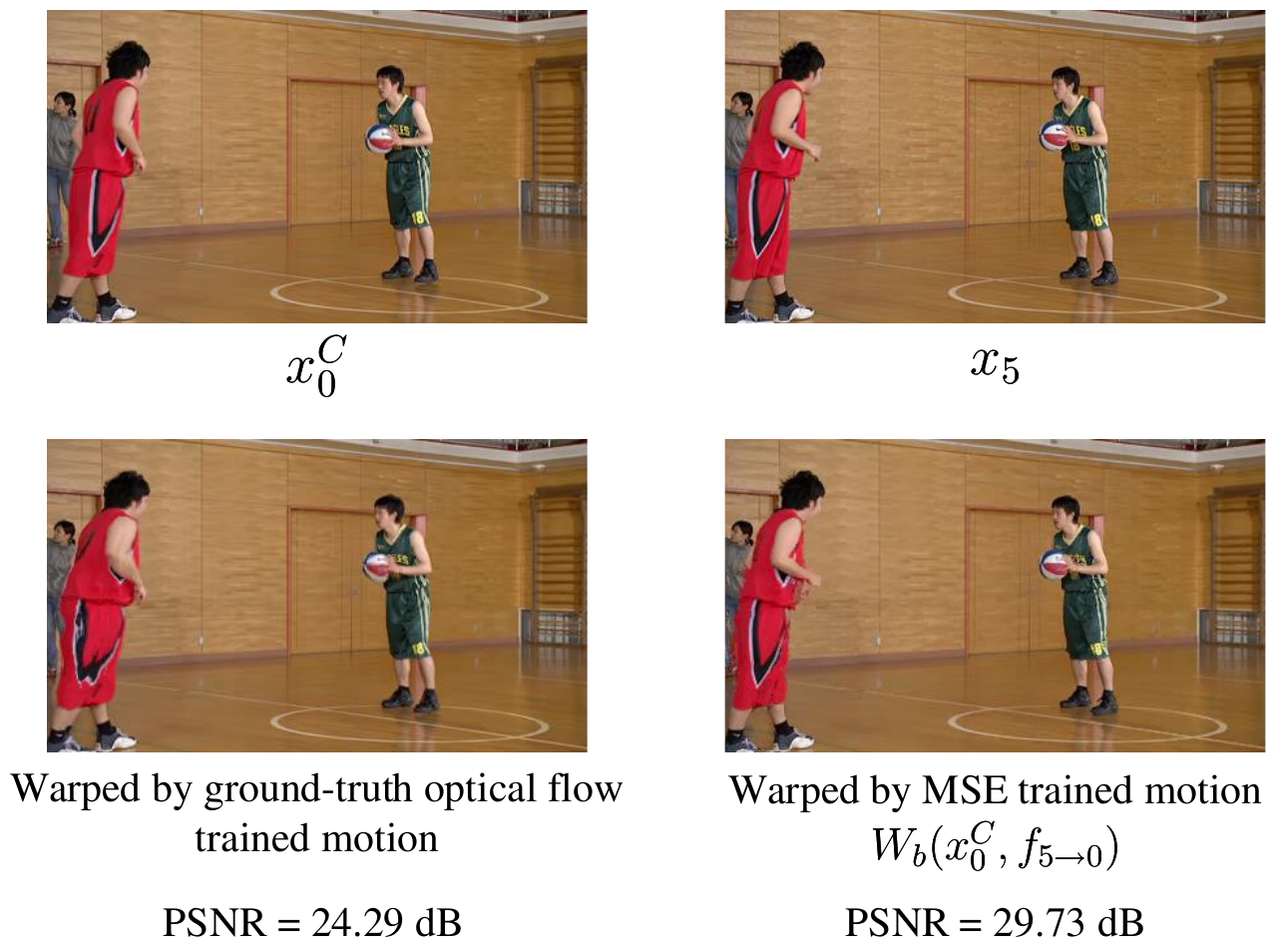}
\end{center}
\vspace{-1em}
\caption{Example frames warped by estimated motions, which are trained by ground-truth optical flow and the MSE loss of \eqref{me}.}\label{fig:me}
\end{figure}

\textbf{ME subnet.} In our approach, we employ a 5-level pyramid network~\cite{ranjan2017optical} for motion compensation, with the same structure and settings as~\cite{ranjan2017optical}. However, \cite{ranjan2017optical} trains the network with the supervision of the ground-truth optical flow, but in our approach, we pre-train the ME subnet by minimizing the MSE between the warped frame and target frame. That is, we use the loss function of
\begin{equation}
    L_\text{ME} = D(x_5, W_b(x_0^C, f_{5\rightarrow 0})) + D(x_5, W_b(x_{10}^C, f_{5\rightarrow 10})) \label{me}
\end{equation}
to initialize our ME subnet before jointly optimizing the whole BDDC network by (12) in our paper. Figure~\ref{fig:me} shows the example frames warped by estimated motions, which are trained by ground-truth optical flow
and the MSE loss of \eqref{me}, and we also show the PSNR between the warped and target frames. It can be seen that the MSE optimized motion is able to reach higher PSNR for the warped frame, thus leading to better motion compensation.

\begin{figure*}[!t]
\begin{center}
\includegraphics[width=.8\linewidth]{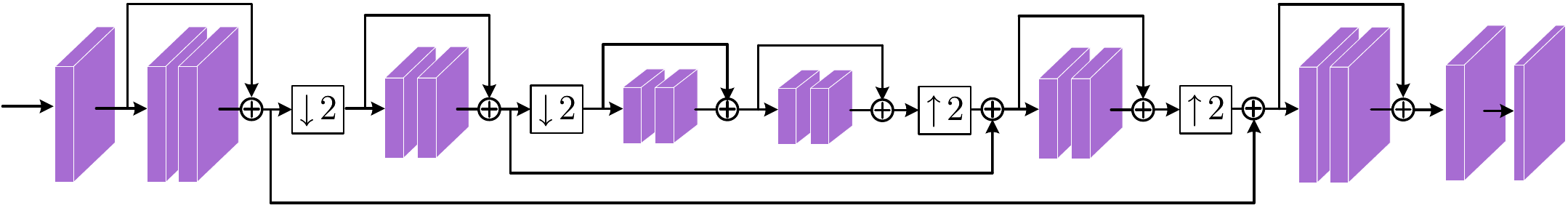}
\end{center}
\caption{Architecture of the MP subnet~\cite{lu2019dvc}.}\label{fig:mp}
\end{figure*}

\begin{figure*}[!t]
\hsize=\textwidth
\centering
\includegraphics[width=.9\linewidth]{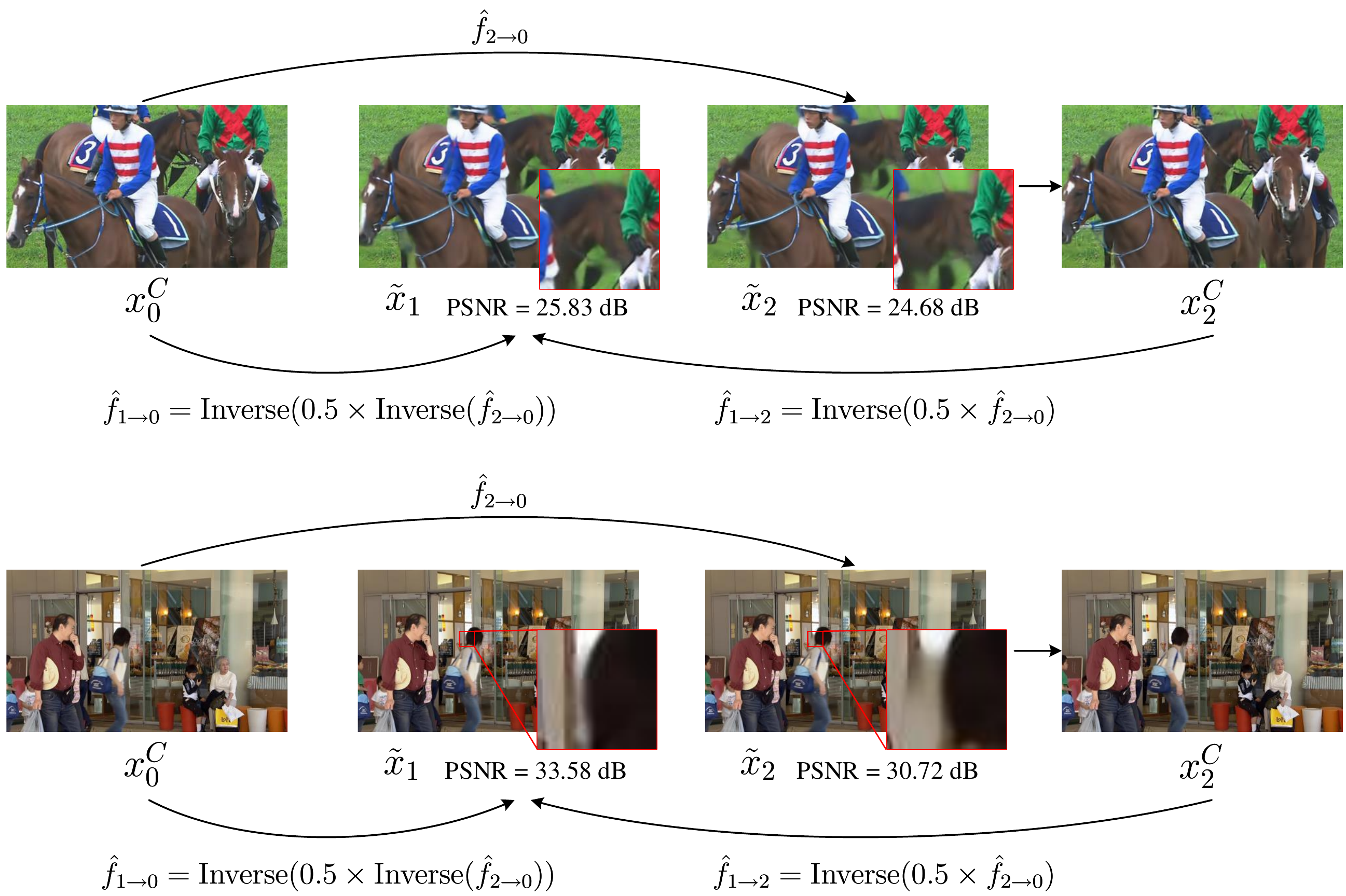}
\caption{Example frames after motion compensation in our SMDC network at $\lambda = 1024$.}
\label{fig:sm}
\end{figure*}

\textbf{MC and RC subnets.}  We follow \cite{balle2017end, balle2018variational} to use the CNN-based auto-encoders in our MC and RC subnets, and they have the same structure in our approach. The detailed parameters are listed in Tables~\ref{tab:encoder} and \ref{tab:decoder}, in which GDN denotes the generalized divisive normalization~\cite{balle2017end} and IGDN is the inverse GDN~\cite{balle2017end}.

\begin{table}[!h]
    \centering

    \caption{The encoder layers in the MC and RC subnets.}
    \begin{tabular}{|c|c|c|c|c|c|}
    \hline
       Layer  & Conv 1 & Conv 2 & Conv 3 & Conv 4 \\
      \hline
       Filter number &  128 & 128 & 128 & 128 \\
       \hline
       Filter size & $5\times 5$ &$5\times 5$  &  $5\times 5$ & $5\times 5$  \\
       \hline
       Activation & GDN & GDN & GDN & - \\
       \hline
       Down-sampling & 2 & 2 & 2 & 2 \\
       \hline
    \end{tabular}
    % \vspace{-1em}
    \label{tab:encoder}
\end{table}

\begin{table}[!h]
    \centering

     \caption{The decoder layers in the MC and RC subnets.}
    \begin{tabular}{|c|c|c|c|c|c|}
    \hline
       Layer  & Conv 1 & Conv 2 & Conv 3 & Conv 4 \\
      \hline
       Filter number &  128 & 128 & 128 & 3 \\
       \hline
       Filter size & $5\times 5$ &$5\times 5$  &  $5\times 5$ & $5\times 5$  \\
       \hline
       Activation & IGDN & IGDN & IGDN & - \\
       \hline
       Up-sampling & 2 & 2 & 2 & 2 \\
       \hline
    \end{tabular}

    \label{tab:decoder}
\end{table}

\textbf{MP subnet.} We use the motion compensation network in \cite{lu2019dvc} as our MP subnet, which is illustrated in Figure~\ref{fig:mp}. In the MP subnet, all convolutional layers use a filter size of $3\times 3$. The filter numbers of all layers excluding the output layer are 64, and the filter number of the output layer is set to 3. We use ReLU as the activation function for all layers.

In our \textbf{SMDC network}, the subnets of ME, MC, MP and RC have the same architecture as introduced above.

\subsection{The WRQE network}

In the WG subnet of our WRQE network, we set the hidden unit number in the bi-directional LSTM as 256, and thus the layer $d_1$ has 512 nodes. We use $5\times 5$ convolutional filters in all convolutional layers (the architecture is shown in Figure \ref{fig:enh} of our paper). The filter numbers are all set to 24 before the output layer, and the filter number for the output layer is 3. We use ReLU as the activation function for all convolutional layers.

\section{Additional experiments}

\textbf{Configurations of x264 and x265.} In Figure \ref{fig:results} and Table \ref{tab:bdbr} of our paper, we follow \cite{lu2019dvc} to use the following settings for x264 and x265, respectively:

\begin{flushleft}
\noindent {\small x264: \texttt{ffmpeg -pix\_fmt yuv420p -s WidthxHeight -r Framerate  -i  Name.yuv -vframes Frame -c:v libx264 -preset veryfast -tune zerolatency -crf Quality -g 10 -bf 2 -b\_strategy 0 -sc\_threshold 0 Name.mkv}}

\vspace{0.1em}

\noindent {\small x265: \texttt{ffmpeg -pix\_fmt yuv420p -s WidthxHeight -r Framerate  -i  Name.yuv -vframes Frame -c:v libx265 -preset veryfast -tune zerolatency -x265-params "crf=Quality:keyint=10:verbose=1" Name.mkv}}
\end{flushleft}

In the commands above, we use Quality = 15, 19, 23, 27 for the JCT-VC dataset, and Quality = 11, 15, 19, 23 for UVG videos.

\textbf{Motion estimation in our SMDC network.} Recall that, in our SMDC network, $\tilde{x}_2$ is generated by motion compensation with the compressed motion $\hat{f}_{2\rightarrow 0}$. Then, we estimate the motions of $\hat{f}_{1\rightarrow 0}$ and $\hat{f}_{1\rightarrow 2}$  from $\hat{f}_{2\rightarrow 0}$ using the inverse operation (refer to \eqref{inv1} and \eqref{inv2} in Section~\ref{sf}) to generate $\tilde{x}_1$.

However, as shown in Figure~\ref{fig:sm}, $\tilde{x}_1$ has even higher PSNR than $\tilde{x}_2$. Moreover, Table~\ref{tab:sm} shows the averaged PSNR of $\tilde{x}_1$ and $\tilde{x}_2$ among all videos in the JCT-VC dataset. The results in Table~\ref{tab:sm} also prove that our SMDC network generates $\tilde{x}_1$ with higher PSNR than $\tilde{x}_2$. It is probably because of the bi-directional motion used for $\tilde{x}_1$, and the shorter distance between $\tilde{x}_1$ and the reference frame $x_0^C$. These results validate that our SMDC network accurately estimates multi-frame motions from a single motion map.

In conclusion, the benefits of our SMDC network can be summarized as:

\begin{table}[!t]
    \centering
    \small
     \caption{Average PSNR (dB) of $\tilde{x}_1$ and $\tilde{x}_2$ in our SMDC network.}
    \begin{tabular}{|c|c|c|c|c|c|}
    \hline
& $\lambda=256$ & $\lambda=512$ & $\lambda=1024$ & $\lambda=2048$ \\
      \hline
       PSNR of $\tilde{x}_1$ & 27.67 & 28.65 & 29.43 & 29.92 \\
       \hline
       PSNR of $\tilde{x}_2$ & 26.42 & 27.44 & 28.22 & 28.58  \\
       \hline
    \end{tabular}
 \vspace{-1.5em}
    \label{tab:sm}
\end{table}

(1) As discussed in Section 4.3 of our paper, our SMDC network reduces the bit-rate for motion information, due to compressing a single motion map in SMDC.

(2) Our SMDC network generates $\tilde{x}_1$ with even higher quality than $\tilde{x}_2$, and thus leads to fewer residual between $\tilde{x}_1$ and ${x}_1$ to encode. This facilitates the residual compression subnet to achieve better compression performance.

%\newpage

\begin{figure*}[!t]
\begin{center}
\vspace{-2em}
\includegraphics[width=.8\linewidth]{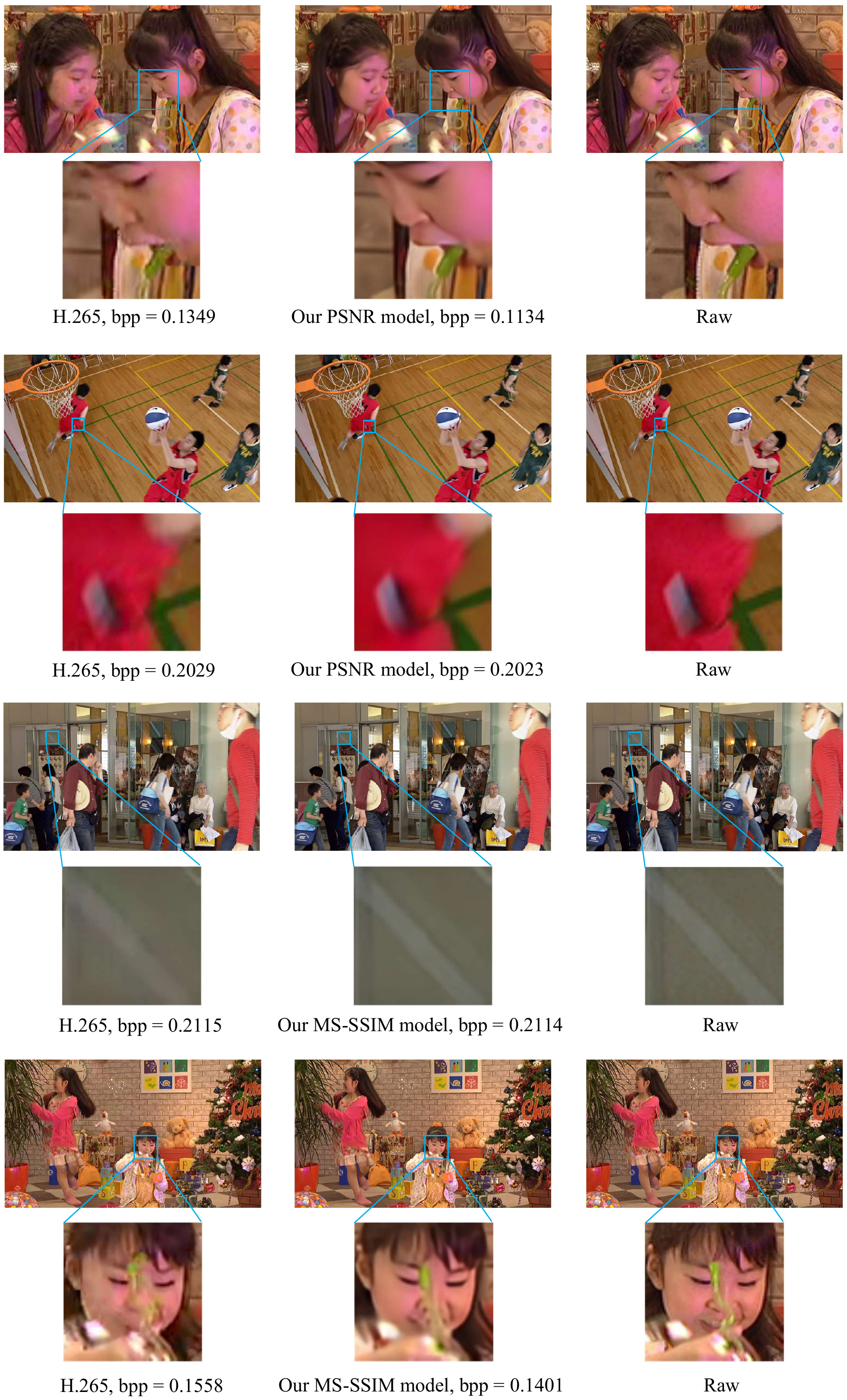}
\end{center}
 \vspace{-1em}
\caption{Visual results of our PSNR and MS-SSIM models in comparison with H.265.}\label{fig:vq}
\vspace{-2em}
\end{figure*}

\textbf{Visual results.} Then, we demonstrate more visual quality results of our PSNR and MS-SSIM models and the latest video coding standard H.265 in Figure~\ref{fig:vq}. The bit-rates in Figure~\ref{fig:vq} are the average values among all frames in each video, and the frames in Figure~\ref{fig:vq} are selected from layer 3, \ie, the lowest quality layer in our approach. It can be seen from Figure~\ref{fig:vq} that both our PSNR and MS-SSIM models have less compression artifacts than H.265, in case that our models consume lower bit-rate. That is, the frames in the lowest quality layer of our approach still achieve better visual quality, when the average bit-rate of the whole video is lower than H.265.

\textbf{Different GOP sizes.} Our method is able to adapt to different GOP sizes, since our BDDC and SMDC networks can be flexibly combined. In Figure \ref{fig:framework}, one more SMDC module can be inserted between the two SMDC modules, enlarging the GOP size to 12. More SMDC modules can be inserted to further enlarge the GOP size. In DVC \cite{lu2019dvc}, GOP = 12 is applied on the UVG dataset. For fairer comparison, we also test our HLVC approach on UVG with GOP = 12. Our PSNR performance on UVG drops to BDBR = $1.53\%$ with the same anchor in Table \ref{tab:bdbr}, but we still outperform Wu~\etal~\cite{wu2018video} and DVC~\cite{lu2019dvc}.

\textbf{Comparison with different configurations of x265.} In our paper, we compare with the ``LDP very fast'' mode of x265.
However, since our HLVC model has a ``hierarchical B'' structure, we further compare our approach with x265 configured with ``b-adapt=0:bframes=9:b-pyramid=1'' instead of ``zerolatency''. The detailed configuration is as follows.

\

\noindent {\small \texttt{ffmpeg -pix\_fmt yuv420p -s WidthxHeight -r Framerate  -i  Name.yuv -vframes Frame -c:v libx265 -preset veryfast/medium -x265-params "b-adapt=0:bframes=9:b-pyramid=1:
crf=Quality:keyint=10:verbose=1" Name.mkv}}

\

With the anchor of x265 ``hierarchical B'', we achieve BDBR = $-9.85\%$ and $-10.55\%$ for the ``medium'' and ``very fast'' modes on MS-SSIM. For PSNR, we do not outperform x265 ``Hierarchical B'' (BDBR = $20.87\%$). Besides, we also compare our approach with the ``LDP medium'' mode of x265, where we obtain BDBR = $-34.45\%$ on MS-SSIM. In terms of PSNR, we are comparable with x265 ``LDP medium'' (BDBR = $-1.08\%$).

\textbf{Comparing ablation results with DVC \cite{lu2019dvc}.} DVC \cite{lu2019dvc} has the IPPP prediction structure, while our approach employs the bi-directional hierarchical structure. To directly compare the performance of these two kinds of frame structure, we calculated the BDBR values of our ablation models with the anchor of DVC \cite{lu2019dvc} on JCT-VC. The results of our ``baseline+HQ'' and ``baseline+HQ+SM'' models are $-5.50\%$ and $-7.37\%$ vs.\ DVC \cite{lu2019dvc}, respectively. It can be seen that our hierarchical model (+HQ) outperforms the IPPP structure of DVC (BDBR$<$0). This validates the effectiveness of our hierarchical layers.

\end{document}